\documentclass[11pt,a4paper]{article}
\usepackage[T1]{fontenc}
\usepackage{graphics}
\usepackage{indentfirst}
\newcommand{\balpha}{\mbox{\boldmath$\alpha$}}
\newcommand{\bomega}{\mbox{\boldmath$\omega$}}
\newcommand{\bmu}{\mbox{\boldmath$\mu$}}
\newcommand{\bbeta}{\mbox{\boldmath$\beta$}}

\begin{document}
\begin{center} 
\huge{\bf Kinematical theory of spinning particles:\\ The interaction Lagrangian
for two spin $1/2$ Dirac particles}
\end{center}
\vspace{0.5cm}

\begin{center}
\large{\bf Mart\'{\i}n Rivas}

\vspace{0.5cm}

\large{{Dpto. de F\'{\i}sica Te\'orica, Universidad del Pa\'{\i}s Vasco-Euskal Herriko Unibertsitatea,\\ 
Apdo.~644, 48080 Bilbao, Spain}\\{http://tp.lc.ehu.es/martin.htm}}
\end{center}

\vspace{0.5cm}

\noindent{\bf Plenary lecture of the Advanced Studies Institute, Symmetries and Spin,\\ Prague
19-26 July 2006}

\vspace{0.5cm}
\noindent{\bf Abstract}

The concept of elementary particle rests on the idea that it is a physical
system with no excited states, so that all possible states of the particle
are just kinematical modifications of any one of them. In this way instead of describing
the particle attributes it amounts to describe the collection of consecutive inertial
observers who describe the particle in the same kinematical state. 
The kinematical
state space of an elementary particle is a homogeneous space of the kinematical group.
By considering the largest homogeneous spaces of both, Galilei and Poincar\'e groups,
it is shown how the spin structure is related to the different degrees of freedom. Finally,
the spacetime symmetry group of a relativistic particle which satisfies Dirac's equation when quantized,
is enlarged to take into account additional symmetries like
spacetime dilations and local rotations. An interaction Lagrangian invariant under this enlarged group
is proposed and the compound system of two Dirac particles is analyzed.

 \section{Introduction} 
We are going to discuss within the kinematical formalism developped by the author, the interaction
Lagrangian for two spin 1/2 Dirac particles. In order to make the contribution selfcontained we shall
describe in section 2 the main theoretical considerations which lead to the definition of a classical
elementary particle. This definition also applies to the spinless point particle 
but in general describe systems which contain
some more variables, such that the usual classical mechanics machinery will produce the definition
of spin in terms of these additional variables. 
The formalism is a variational formalism which is quantized under Feynman's path integral
method. It makes a prediction concerning the electromagnetic dipole structure of particles and antiparticles
in such a way that spin and magnetic moment must have the same relative orientation for matter and antimatter.
Two plausible experiments are proposed to determine this relative orientation. 

In section 3 we discuss the enlargement of the spacetime symmetry group to include also spacetime
dilations and local rotations and the relationship of the new spacetime symmetry group with
the standard model. Finally, in section 4 we describe the interaction Lagrangian for a system of two
interacting Dirac's particles which is obtained under the assumption that the intrinsic structure of an 
elementary particle is not modified by any interaction. 
With this Lagrangian we analyze the interaction
of two particles and the possibility of formation of bound pairs of equal 
charged particles provided
some boundary conditions are fulfilled.

 \section{Kinematical formalism} 
We call the present approach a {\it kinematical formalism} because 
the spacetime symmetry group of the theory, the kinematical group,
not only supplies the symmetries, and therefore the conserved observables
of the system, but also
its group parameters, compact and non compact, will be transformed
into the classical variables we need to describe the elementary systems \cite{Rivasbook}.
Classical elementary particles are localized and orientable systems. By this we mean that
the position of the particle is described by the evolution of a geometrical point where
all the particle charges responsible for its interaction are located, and the orientation
is described by attaching a comoving cartesian frame which describes the particle rotation. 
Nevertheless the position of that point is not the center of mass for a spinning particle,
so that the motion of the charge around the center of mass produces the dipole structure.
For a Dirac particle the charge moves at the speed of light so that it is never at rest 
for any inertial observer and this justifies the coupling of the particle current with the
external potentials.
The particle is characterized by two kinds of classical variables.
Non-compact variables like time $t$, position ${\bf r}$ and 
compact variables like the orientation variables $\balpha$.
The spin $1/2$ structure of elementary particles 
will be related to compact variables and, therefore, 
mathematical theorems on compact groups will play an important role in the
quantum mechanical structure of spin. 

The dynamics is based upon a variational formalism. The initial and final states
of the classical variational formalism, characterized by what will be called {\it kinematical variables}, 
will correspond with the initial and final states
of the quantum dynamical formalism.
We have to find solutions passing through the end points so that we have to give a
protagonism to the kinematical variables.
The Lagrangian must be rewritten in terms of these variables, instead of the independent degrees 
of freedom.

	We also accept the {\it atomistic hypothesis}.
Matter cannot be divided indefinitely. After a finite number of steps we reach a final
and indivisible object, i.e., an elementary particle. 
Real matter does not satisfy the hypothesis of the continuum.
An {\it elementary particle} is a mechanical system without excited states. We can destroy
it but we can never modify its structure. All its possible states are only kinematical 
modifications of any one of them.
If the state of an elementary particle changes, it is always possible to find
another inertial observer who describes the particle in the same state as in the previous instant.

Any elementary system is thus a Lagrangian system whose initial and final states
can be characterized by a finite number of variables. By the above consideration on the
states of an elementary particle
the kinematical variables necessarily span a homogeneous space of
the kinematical group associated to the restricted Relativity Principle.
The variables which characterize the classical states of an elementary particle are thus 
the variables which characterize the possible parameterizations of the kinematical group 
of space-time transformations.

For any generalized Lagrangian system $L(t, q_i, q_i^{(1)},\ldots, q_i^{(k-1)}, q_i^{(k)})$, 
of $n$ degrees of freedom $q_i$, and whose Lagrangian depends on the derivatives up to a finite order
$k$-th, the kinematical variables are 
\[
 x\equiv t, q_i, q_i^{(1)},\ldots, q_i^{(k-1)}.
\]
If, instead of time $t$ we use an arbitrary evolution parameter $\tau$
\[
\int_{t_1}^{t_2}L(t, q_i,q_i^{(1)},\ldots,q_i^{(k-1)}\;\,,q_i^{(k)})dt=
\int_{\tau_1}^{\tau_2}L(x_j\,,\dot{x}_j)d\tau, 
\]
where $\dot{x}=dx/d\tau$, $q^{(1)}=dq/dt=\dot{q}/\dot{t}$, $q^{(2)}=dq^{(1)}/dt=\dot{q^{(1)}}/\dot{t}$, etc., 
$dt=\dot{t}d\tau$, and the Lagrangian becomes a homogeneous function
of first degree of the $\tau-$derivatives of the kinematical variables $\dot{x}_j$,
 \begin{equation}
L(x,\dot{x})=\frac{\partial L}{\partial\dot{x}_j}\dot{x}_j= F_j(x,\dot{x})\dot{x}_j.
 \label{eq:F}
 \end{equation}
Noether's theorem implies that the different constants of the motion can be expressed
only in terms of the functions $F_j$ and their time derivatives.

The generalized coordinates are $ q_i,q_i^{(1)},\ldots,q_i^{(k-1)}$, and their
canonical conjugate momenta
\[
p_{is}=\sum_{r=0}^{k-s}(-1)^r\frac{d^{r}}{dt^r}F_{(r+s-1)n+i},\quad i=1,\ldots,n,\quad s=1,\ldots,k
\]
and the Hamiltonian is
\[
H=p_{is}q_i^{(s)}-L
\]where
$F_j(x,\dot{x})$ are given in (\ref{eq:F}), and thus the phase space is of dimension $2kn$.

If $G$ is a $r-$parameter symmetry group of parameters $g^\alpha$, $\alpha=1,\ldots,r$, 
which leaves invariant the Lagrangian and transforms
infinitesimally the kinematical variables in the form
\[
x'_j=x_j+M_{j\alpha}(x)\delta g^\alpha,\quad j=0,1,\ldots, kn,\quad \alpha=1,\ldots,r,
\]
the $r$ conserved Noether's observables are given by
\[
N_\alpha= HM_{0\alpha}(x)-p_{is}M_{\{(s-1)n+i\}\alpha}(x),\quad i=1,\ldots,n,\quad s=1,\ldots k.
\]
The advantage of this formulation is that we can obtain general expressions for the conserved quantities
in terms of the above $F_i(x,\dot{x})$ functions, and their time derivatives, which are homogeneous functions of zero-th degree
of the variables $\dot{x}_i$ and of the way the kinematical variables transform $M_{j\alpha}(x)$.

The quantization
leads to the following results:\cite{Rivas2}
\begin{enumerate}
\item{If $x$ are the kinematical variables of the variational approach, 
Feynman's kernel $K(x_1,x_2)$ which describes the probability amplitude for the evolution of the system
between the initial point $x_1$ to the final point $x_2$, is only a function (more properly a distribution)
of the end point kinematical variables.}
\item{ The wave function of the quantized system
is a complex squared integrable function of these variables $\Phi(x)$, 
with respect to some suitable invariant measure
over the kinematical space.}
\item{The wave function transforms under the kinematical group
\[
\Phi'(x)=U(g)\Phi(x)=\Phi(g^{-1}x)e^{-i\alpha(g^{-1};x)/\hbar}
\]
with a projective unitary irreducible representation of the kinematical group.}
\item{If $G$ is a symmetry group of parameters $g^\alpha$, which transform infinitesimally 
the kinematical variables in the form:
\[
x'_j=x_j+M_{j\alpha}(x)\delta g^\alpha,
\]
the representation of the generators is given by the self-adjoint operators $(\hbar=1)$
\[
X_\alpha=-iM_{j\alpha}(x)\frac{\partial }{\partial x_j}.
\]}

\item{Quantum theory is not a hidden variable theory because it describes the quantum states
in terms of a wave function which depends on exactly the same classical variables as the classical mechanics does.
\[
x\quad\rightarrow\quad \Phi(x)
\]
and thus no classical information is lost in the proccess of quantization.}

\end{enumerate}

In our model of the Dirac particle, the kinematical variables are time $t$, the position of the charge ${\bf r}$,
the velocity of the charge ${\bf u}$ with the constraint $u=c$, and, finally the orientation
of the particle $\balpha$ described by a suitable parameterization of the rotation group and which describes
the orientation of a local body frame attached to the point ${\bf r}$. 
According to the homogeneity condition (\ref{eq:F}) the general form of the Lagrangian is
 \begin{equation}
L=T\dot{t}+{\bf R}\cdot{\dot{\bf r}}+{\bf U}\cdot{\dot{\bf u}}+{\bf W}\cdot{\bomega},
 \label{eqLag}
 \end{equation}
where $T={\partial L}/{\partial\dot{t}}$, ${\bf R}={\partial L}/{\partial\dot{\bf r}}$, ${\bf U}={\partial L}/{\partial\dot{\bf u}}$
and  ${\bf W}={\partial L}/{\partial{\bomega}}$. All conserved quantities obtained by applying Noether's theorem, 
will be expressed in terms of these functions and their time derivatives.

We see that the Lagrangian depends up to the acceleration of the point ${\bf r}$ and therefore
dynamical equations for the motion of this point are fourth order differentail equations \cite{dyn}.
The general expressions of the ten conserved Nother's observables are
\[
H=-T+{\bf u}\cdot\frac{d{\bf U}}{dt},\quad {\bf P}={\bf R}-\frac{d{\bf U}}{dt},
\]
\[
{\bf K}=\frac{1}{c^2}H{\bf r}-{\bf P}t+\frac{1}{c^2}{\bf u}\times{\bf S},\quad {\bf J}={\bf r}\times{\bf P}+{\bf u}\times{\bf U}+{\bf W}={\bf L}+{\bf S},
\]
which are called respectively, energy, linear momentum, kinematical momentum and angular momentum.
We see the twofold structure of the spin observable which depends on the term 
${\bf Z}={\bf u}\times{\bf U}$ or zitterbewegung part and the rotational part ${\bf W}$.
The center of mass position is defined by
\[
{\bf q}={\bf r}-\frac{1}{H}{\bf S}\times{\bf u},
\]  
which is always different from ${\bf r}$, whenever the spin is not zero. The time derivative of the conserved
${\bf J}$ brings for the spin the dynamical equation
 \begin{equation}
\frac{d{\bf S}}{dt}={\bf P}\times{\bf u},
 \label{sdyn}
 \end{equation}
which is not conserved because ${\bf P}$ and ${\bf u}$ are not collinear vectors. 
This is the same dynamical equation satisfied by Dirac's spin operator in the quantum case.
We thus see that our spin observable ${\bf S}$ represents the angular momentum of the particle
with respect to the charge position ${\bf r}$ and not with respect to the center of mass ${\bf q}$. This is one of the
reasons why this magnitude is not a conserved one even for a free particle. Is is only conserved in the centre
of mass frame, where ${\bf P}=0$.

The wave function becomes a complex squared integrable function 
defined on the kinematical space $\Phi(t,{\bf r},{\bf u},{\balpha})$.
The Poincar\'e group unitary realization over the corresponding Hilbert space 
has the usual selfadjoint generators. They are represented by the differential operators,
with respect to the kinematical variables, in three-vector form ($\hbar=c=1$):
\[
H=i{\partial }/{\partial t},\quad {\bf P}=-i\nabla,
\]
\[
 {\bf K}=i{\bf r}{\partial }/{\partial t}+it{\nabla}+{\bf u}\times{\bf S},\quad {\bf J}=-i{\bf r}\times\nabla+{\bf S}={\bf L}+{\bf S}.
\]
The spin operator ${\bf S}$ is given by
\[
S_i=-i\epsilon_{ijk}u_j{\partial }/{\partial u_k}+W_i,\quad\hbox{\rm or}
\quad{\bf S}=-i{\bf u}\times{\nabla_u}+{\bf W}={\bf Z}+{\bf W}.
\]
$\nabla_u$ is the gradient operator with respect to the $u_i$ variables and the ${\bf W}$ operator involves 
differential operators with respect to the orientation variables. Its structure depends 
on the selection of the variables which represent the orientation and which correspond 
to the different parameterizations
of the rotation group. In the normal or canonical parameterization of the rotation group, every rotation
is characterized by a three vector $\balpha=\alpha{\bf n}$, where ${\bf n}$ is a unit vector along the rotation axis
and $\alpha$ the clokwise rotated angle. If we represent the unit vector ${\bf n}$ by the usual polar 
and azimuthal angles $(\theta,\phi)$,
$\theta\in[0,\pi]$ and $\phi\in[0,2\pi]$,
then every rotation is parameterized by the three dimensionless variables $(\alpha,\theta,\phi)$. In this parameterization
the $W_i$ spin operators become:
 \begin{eqnarray}
W_1&=&{1\over 2i}\left[2\sin\theta\,\cos\phi\,{\partial\over\partial\alpha}+ 
\left({\cos\theta\,\cos\phi\over\tan(\alpha/2)}-\sin\phi\right){\partial\over\partial\theta}\,-\right.\nonumber\\
&&\left.\left({\sin\phi\over\tan(\alpha/2)\sin\theta}+{\cos\theta\, 
\cos\phi\over\sin\theta}\right)\,{\partial\over\partial\phi}\right],
 \label{eq:Y1}
 \end{eqnarray}
 \begin{eqnarray}
W_2&=&{1\over 2i}\left[2\sin\theta\,\sin\phi\,{\partial\over\partial\alpha}+ 
\left({\cos\theta\,\sin\phi\over\tan(\alpha/2)}+\cos\phi\right){\partial\over\partial\theta}\,-\right.\nonumber\\
&&\left.\left({\cos\theta\,\sin\phi\over\sin\theta}-{\cos\phi\over\tan(\alpha/2)\sin\theta}\right) 
{\partial\over\partial\phi}\right],
 \label{eq:Y2}
 \end{eqnarray}
 \begin{eqnarray}
W_3={1\over 2i}\left[2\cos\theta\,{\partial\over\partial\alpha}-
{\sin\theta\over\tan(\alpha/2)}{\partial\over\partial\theta}+ 
{\partial\over\partial\phi}\right],
 \label{eq:Y3}
 \end{eqnarray}
The $Z_i=-i{\bf u}\times{\nabla_u}$ operators are expressed in terms of the direction $\widetilde{\theta},\widetilde{\phi}$
of the velocity ${\bf u}$ in the form:
\[
Z_1=i\sin{\widetilde\phi}\frac{\partial }{\partial\widetilde\theta}+i\frac{\cos{\widetilde\theta}}{\sin{\widetilde\theta}}\cos{\widetilde\phi}\frac{\partial }{\partial\widetilde\phi},\quad
Z_2=-i\cos{\widetilde\phi}\frac{\partial }{\partial\widetilde\theta}+i\frac{\cos{\widetilde\theta}}{\sin{\widetilde\theta}}\sin{\widetilde\phi}\frac{\partial }{\partial\widetilde\phi},\]
\[
Z_3=-i\frac{\partial }{\partial\widetilde\phi}.
\]
and therefore is eigenvectors are the spherical harmonics $Y_l^m(\widetilde{\theta},\widetilde{\phi})$.
They satisfy the commutation relations
\[
[L_j, L_k]=i\epsilon_{jkl} L_l,\quad [Z_j, Z_k]=i\epsilon_{jkl} Z_l,\quad[W_j, W_k]=i\epsilon_{jkl} W_l,\]
\[
[L_i,Z_j]=[L_i,W_k]=[Z_i,W_k]=0.
\]
and thus
\[
[J_j, J_k]=i\epsilon_{jkl} J_l,\quad [J_i,H]=[J_i,D]=0,\quad [J_j, P_k]=i\epsilon_{jkl} P_l.
\]
We thus clearly see that the structure of the quantum mechanical spin operator ${\bf S}$ only takes derivatives 
with respect to the compact kinematical variables $\widetilde\theta,\widetilde\phi,\alpha,\theta,\phi$.

The classical expression that leads to Dirac equation when quantizing the system comes from 
the conserved kinematical momentum ${\bf K}$.
\[
{\bf K}=\frac{H}{c^2}{\bf r}-{\bf P}t-{\bf S}\times\frac{{\bf u}}{c^2},
\quad\Rightarrow\quad
\frac{d{\bf K}}{dt}=0=\frac{H}{c^2}{\bf u}-{\bf P}-\frac{d}{dt}\left({\bf S}\times\frac{{\bf u}}{c^2}\right)
\]a subsequent scalar product with the velocity leads to
 \begin{equation}
H={\bf P}\cdot{\bf u}+\frac{1}{c^2}{\bf S}\cdot\left(\frac{d{\bf u}}{dt}\times{\bf u}\right).
 \label{eq:Dir}
 \end{equation}
This is a linear relationship between $H$ and ${\bf P}$, where the velocity ${\bf u}$ 
should be replaced by Dirac's velocity operator $c\balpha$
and the last term corresponds to $\beta mc^2$ in terms of Dirac's $\beta$ matrix.
In the center of mass frame the absolute
value of the acceleration is $c^2/R$, so that taking into account the value of $R$ we get that this term
reduces to $\pm mc^2$, the positive value for the particle and the negative one for the antiparticle.

Taking into account that the spin dynamical equation, even for a free particle is (\ref{sdyn}),
we arrive to the classical expression of the spin in terms of the internal motion
 \begin{equation}
{\bf S}=\frac{H-{\bf P}\cdot{\bf u}}{|{d{\bf u}}/{dt}|^2}\left(\frac{d{\bf u}}{dt}\times{\bf u}\right),
 \label{eq:spin}
 \end{equation}
so that classical Dirac's spin observable is always orthogonal to the velocity and acceleration of the charge.

\begin{figure}
\begin{center}
\includegraphics{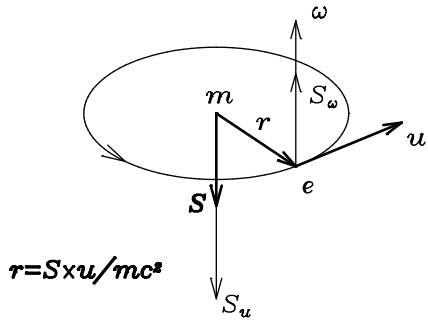}
\caption{Motion of the charge of the electron in the center of mass frame. The magnetic moment is produced by the
motion of the charge and the particle also has an oscillating electric dipole moment of value ${\bf d}=e{\bf r}$.}
\label{fig:el}
\end{center}
\end{figure}

For the center of mass observer ${\bf K}={\bf P}=0$, the spin is a constant 
of the motion, $H=mc^2$ and thus 
 \[
{\bf K}=\frac{H}{c^2}{\bf r}-{\bf P}t-\frac{1}{c^2}{\bf S}\times{\bf u}=0,\quad\Rightarrow\quad 
{\bf r}=\frac{1}{mc^2}{\bf S}\times{\bf u},
\]
so that point ${\bf r}$ is moving in circles, 
at the speed of light, on a plane orthogonal to the constant vector ${\bf S}$ as shown in Fig.{\ref{fig:el}}.
Classical mechanics does not restrict the value of the constant spin $S$ which 
can be any positive real number.
Its true value will be uniquely fixed after quantization.
The radius of this circle is $R=S/mc$ and the angular velocity of this internal 
motion or {\it zitterbewegung} is $\omega=mc^2/S$. 

In fact, in the center of mass frame, it is a 
system of three degrees of freedom. The coordinates $x$ and $y$ of the
point ${\bf r}$ and the phase $\alpha$ of the rotation 
of the body frame. This phase is the same as the phase 
of the orbital motion. The
motion is at a constant velocity $c$, then the system
is reduced to a single degree of freedom system.
It is a one-dimensional harmonic oscillator of frequency $\omega=mc^2/S$,
without excited states. The ground state energy of this system when quantized, 
is $\hbar\omega/2=mc^2$ which implies that the classical parameter $S=\hbar/2$.

The negative energy particle corresponds to the time reversed motion with the same spin ${\bf S}$.
But the formalism does not fix the sign of the charge.

\subsection{Chirality and PCT invariance}

\begin{figure}[t]
\begin{center}
\includegraphics{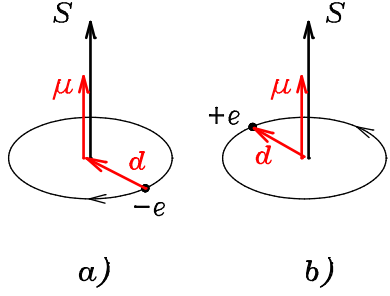}
\caption{Matter is lefthanded and antimatter is righthanded, as far as the motion of the charge
is concerned once the spin direction is fixed.
Particle and antiparticle have the same mass and spin and also the same
electric and magnetic dipole with the same relative orientation with respect to the spin. The antiparticle
in $b)$ is the $PCT$ transformed of the particle $a)$}
\label{fig:chir}
\end{center}
\end{figure}

Matter is lefthanded and antimatter is righthanded when we consider the motion of the charge
with respect to the spin direction. Matter moves counterclockwise when looking along 
the direction of spin. Antimatter moves in the opposite direction.
If we assume that the positive energy particle
is of negative charge we obtain the dipole structure of the particle and antiparticle as depicted in Fig.{\ref{fig:chir}}.
If the particle is of positive charge we obtain the opposite relative orientation.
This indefiniteness in the sign of the charge of matter is also present in Dirac's formalism.
This prediction is consistent with the known structures formed by a particle 
and the corresponding antiparticle.
The positronium is an unstable bound state between an electron and a positron. The ground state,
the singlet parapositronium state,
is of zero spin and zero magnetic moment, thus justifying that particle 
and antiparticle have the same
relative orientation between the spin and magnetic moment. 
In the case of the bound system $q\bar{q}$ betweeen two quarks we have the $\pi^0$ which is a zero 
spin and zero magnetic moment particle which is represented as a 
linear combination of $u\bar{u}$, and $d\bar{d}$ quarks and antiquarks.

Nevertheless, the theory does not predict the relative orientation because does not fix the sign of the charge
of the particle, as in Dirac's formalism. This has to be measured experimentaly. In our opinion this feature
has not been measured and we propose two plausible experiments to establish their relative orientation.

\subsection{Two plausible experiments}
We propose an experiment for the measurement of the relative orientation 
between ${\bf S}$ and ${\bmu}$ for an electron bound in a Rb atom.
Rb$^{87}$ atom has a $5s$ outer electron. Its nucleus has spin $3/2$ and the ground state
of the atom corresponds to a system of total spin 1, with the spin of the outer electron
opposite to that of the nucleus.
The magnetic moment of the atom is basically the magnetic moment of the outer electron.
Ultracold atoms in an external magnetic field will orient their magnetic moments
along the direction of the field.

If we send in this direction circularly polarized photons of such an energy to produce
the hyperfine transition which flips the spin of the electron to the opposite orientation,
and therefore the atom goes into a total spin 2 state, then
only those photons with a spin opposite to the outer electron
will be absorbed.

Another experiment is the measurement of the preccession direction of the spin 
of $e^+$ and $e^-$ and of $\mu^+$ and $\mu^-$ in a storage ring.
If $e^+$ and $e^-$ and $\mu^+$ and $\mu^-$ have the same relative orientation between spin and 
magnetic moment, then the torque and thus the preccession will be the same. 
\[
\bmu\times{\bf B}=\frac{d{\bf S}}{dt}
\]
Nevertheless, if we inject into the accelerator particles and antiparticles with the spin up,
and because the magnetic field of the ring has to be reversed  for the antiparticle, 
then the preccession direction of both beams
will be opposite to each other. If it is possible to detect the precession direction 
this will confirm the prediction and also the relative orientation between spin and magnetic moment. 

\section{More spacetime symmetries}

The wave function is a function of the kinematical variables
\[
\Phi(t,{\bf r};\widetilde{\theta},\widetilde{\phi},\alpha,\theta,\phi)=
\sum_{i=1}^{i=4}\psi_i(t,{\bf r})\Phi_i(\widetilde{\theta},\widetilde{\phi},\alpha,\theta,\phi)
\] 
If the system has spin $S$ and mass $m$ we can define a length scale $R=S/mc$ 
and a time scale $T=S/mc^2$, so that the 9 kinematical variables of a Dirac particle
can be taken dimensionless from the classical point of view. The length scale is the radius 
of the internal motion and the time scale the time taken during a complete turn of the charge.

This means that this system, in addition to the Poincar\'e group, 
it also has as a symmetry group the space-time 
dilations which do not change the speed of light
\[
t'=e^\lambda t,\quad {\bf r}'=e^\lambda{\bf r} ,\quad{\bf u}'={\bf u},\quad \balpha'=\balpha.
\]
the generator of the unitary representation of this $U(1)$ group is
\[
D=it\frac{\partial }{\partial t}+i{\bf r}\cdot\nabla=tH-{\bf r}\cdot{\bf P}.
\]
The enlarged group is the Weyl group, denoted here by ${\cal W}$.

The Poincar\'e group ${\cal P}$ has two Casimir operators expressed in terms of the four-momentum
$P_\mu$ and the Pauli-Ljubansky operator $W_\mu$, in the form
\[
C_1=P^\mu P_\mu=H^2-{\bf P}^2=m^2,\quad C_2=-W^\mu W_\mu=m^2 s(s+1)\hbar^2.
\]
The enlarged Weyl group ${\cal W}$ has only one Casimir operator which for massive systems ($C_1$ is invertible) 
is reduced to
\[
C=\frac{C_2}{C_1}=s(s+1)\hbar^2.
\]
The {\bf spin} is the only intrinsic property of this elementary particle.

The rotation group acts on the kinematical variables in the way:
\[
t'=t,\quad {\bf r}'=R(\bmu){\bf r},\quad {\bf u}'=R(\bmu){\bf u},\quad
R(\balpha')=R(\bmu)R(\balpha).
\]
But the orientation variables $\balpha$ are arbitrary, so that we can also have another local 
rotation body frame transformations
\[
t'=t,\quad {\bf r}'={\bf r},\quad {\bf u}'={\bf u},\quad
R(\balpha')=R(\bbeta)R(\balpha).
\]
This corresponds to the active rotation of the body frame ${\bf e}_i$, $i=1,2,3$. The generators of these rotations are the spin projections on the body frame, {\it i.e.}
the $T_i={\bf e}_i\cdot{\bf W}$ operators.
These operators commute with the whole ${\cal W}$ group. So that the new space-time group is 
${\cal W}\otimes SO(3)$ and becomes in the quantum case
${\cal W}\otimes SU(2)$.
It has two Casimir operators ${\bf S}^2$ and ${\bf T}^2$. But because ${\bf T}^2={\bf W}^2$
the eigenvalues of ${\bf T}^2$ are only $1/2$.

The $T_i$ operators are differential operators with respect to the compact orientation variables $\alpha,\theta,\phi$,
and are given by
 \begin{eqnarray}
T_1&=&{-i\over 2}\left[2\sin\theta\,\cos\phi\,{\partial\over\partial\alpha}+ 
\left({\cos\theta\,\cos\phi\over\tan(\alpha/2)}+\sin\phi\right){\partial\over\partial\theta}\;-\nonumber\right.\\
&&\left.\left({\sin\phi\over\tan(\alpha/2)\sin\theta}-{\cos\theta\, 
\cos\phi\over\sin\theta}\right)\,{\partial\over\partial\phi}\right],
 \label{eq:T1}
 \end{eqnarray}
 \begin{eqnarray}
T_2&=&{-i\over 2}\left[2\sin\theta\,\sin\phi\,{\partial\over\partial\alpha}+ 
\left({\cos\theta\,\sin\phi\over\tan(\alpha/2)}-\cos\phi\right){\partial\over\partial\theta}\;-\nonumber\right.\\
&&\left.\left(-{\cos\theta\,\sin\phi\over\sin\theta}-{\cos\phi\over\tan(\alpha/2)\sin\theta}\right) 
{\partial\over\partial\phi}\right],
 \label{eq:T2}
 \end{eqnarray}
 \begin{equation}
T_3={-i\over 2}\left[2\cos\theta\,{\partial\over\partial\alpha}-
{\sin\theta\over\tan(\alpha/2)}{\partial\over\partial\theta}- 
{\partial\over\partial\phi}\right].
 \label{eq:T3}
 \end{equation}  
The $T_i$ are related to the $W_i$ by changing $\alpha$ into $-\alpha$,
and satisfy the commutation relations
\[
[T_i,T_j]=-i\epsilon_{ijk}T_k
\]
wich corresponds to an active rotation.\cite{Rivasst}

Because the spin has the form ${\bf S}={\bf Z}+{\bf W}$ and quantizes with $s=1/2$ while 
${\bf W}$ quantizes with $w=1/2$, the zitterbewegung part ${\bf Z}$ quantizes with $z=0$ or $z=1$.

There are thus two kinds of Dirac's particles according to the two possible ${\bf Z}$ eigenvalues.

\subsection{Standard model?}

If we interpret the new $SU(2)$ local rotation group as representing {\bf isospin}
and the zitterbewegung angular momentum ${\bf Z}$ as representing {\bf color}, the above Dirac particle
with the ${\cal W}\otimes SU(2)$ as its space-time symmetry group is:
\begin{itemize}
\item{A particle of spin 1/2
and of isospin 1/2 of arbitrary nonvanishing mass and arbitrary charge.}
\item{it can be a colourless particle $z=0$ (lepton?), or
a coloured one $z=1$ (quark?). The last one can also be in one of the three $Z_3$ states $1,0,-1$
but the $Z_3$ is unobservable because the four basic $\Phi_i(\widetilde{\theta},\widetilde{\phi},\alpha,\theta,\phi)$
spinors are eigenvectors of ${\bf Z}^2$, ${\bf S}^2$ and ${\bf T}^2$ but not of $Z_3$ for $z=1$ case.}
\end{itemize}
The $\Phi_i$ spinors for the $z=0$ case are in the notation $|0;s_3,t_3>$
 \begin{eqnarray}
\Phi_1&=&|0;1/2,-1/2>=i\sqrt{2}\sin(\alpha/2)\sin\theta e^{i\phi},\label{eq:Fi1}\\
\Phi_2&=&|0;-1/2,-1/2>=\sqrt{2}\left(\cos(\alpha/2)-i\cos\theta\sin(\alpha/2)\right)\\
\Phi_3&=&|0;1/2,1/2>=-\sqrt{2}\left(\cos(\alpha/2)+i\cos\theta\sin(\alpha/2)\right),\\
\Phi_4&=&|0;-1/2,1/2>=-i\sqrt{2}\sin(\alpha/2)\sin\theta e^{-i\phi}.\label{eq:Fi4}
 \end{eqnarray}
They form an orthonormal set with respect to the normalized invariant measure defined on $SU(2)$
\[
dg(\alpha,\theta,\phi)=\frac{1}{4\pi^2}\sin^2(\alpha/2)\sin\theta\,d\alpha\,d\theta\,d\phi,\]
\[
 \alpha\in[0,2\pi],\quad \theta\in[0,\pi],\quad \phi\in[0,2\pi].
\]
\[
\int_{SU(2)} \,dg(\alpha,\theta,\phi)=1.
\]
For the $z=1$ case in the notation $|1;s_3,t_3>$, we represnt them by $\Psi_i$ and they are
 \begin{eqnarray}
{\Psi}_1&=&|1;1/2,1/2>=\frac{1}{\sqrt{3}}\left(Y_1^0(\widetilde{\theta},\widetilde{\phi})\Phi_1-\sqrt{2}Y_1^1(\widetilde{\theta},\widetilde{\phi})\Phi_2\right),\label{eq:Fit1}\\
{\Psi}_2&=&|1;-1/2,1/2>=\frac{1}{\sqrt{3}}\left(-Y_1^0(\widetilde{\theta},\widetilde{\phi})\Phi_2+\sqrt{2}Y_1^{-1}(\widetilde{\theta},\widetilde{\phi})\Phi_1\right),\\
{\Psi}_3&=&|1;1/2,-1/2>=\frac{1}{\sqrt{3}}\left(Y_1^0(\widetilde{\theta},\widetilde{\phi})\Phi_3-\sqrt{2}Y_1^1(\widetilde{\theta},\widetilde{\phi})\Phi_4\right),\\
{\Psi}_4&=&|1;-1/2,-1/2>=\frac{1}{\sqrt{3}}\left(-Y_1^0(\widetilde{\theta},\widetilde{\phi})\Phi_4+\sqrt{2}Y_1^{-1}(\widetilde{\theta},\widetilde{\phi})\Phi_3\right).\label{eq:Fit4}
 \end{eqnarray}
where the $\Phi_i$ are the same as the ones in (\ref{eq:Fi1}-\ref{eq:Fi4}) and the spherical harmonics $Y^i_1(\widetilde{\theta},\widetilde{\phi})$ are
\begin{equation}
Y_1^1=-\sqrt{\frac{3}{8\pi}}\sin(\widetilde{\theta})e^{i\widetilde{\phi}},\;\;
Y_1^0=\sqrt{\frac{3}{4\pi}}\cos(\widetilde{\theta}),\;\;
Y_1^{-1}=\sqrt{\frac{3}{8\pi}}\sin(\widetilde{\theta})e^{-i\widetilde{\phi}}.
\label{esfhar}
\end{equation}
The four spinors ${\Psi}_i$ are orthonormal with respect to the invariant measure
\[
dg(\alpha,\theta,\phi\,;\widetilde{\theta},\widetilde{\phi})=\frac{1}{4\pi^2}\sin^2(\alpha/2)\sin\theta\sin\widetilde\theta\,d\alpha\,d\theta\,d\phi\,d\widetilde{\theta}d\widetilde{\phi}\]
\[
 \alpha\in[0,2\pi],\quad \widetilde{\theta},\theta\in[0,\pi],\quad \widetilde{\phi},\phi\in[0,2\pi].
\]
We clearly see that the $\Psi_i$ are eigenvectors of ${\bf Z}^2$ with eigenvalue $z=1$, but they 
are not eigenvectors of $Z_3$ because they are linear combinations of spherical harmonics of different
$z_3$ values.

\subsection{Larger kinematical space}
Once we have a larger kinematical group we can have a larger kinematical space.
The new kinematical variable $\beta$ associated to the spacetime dilation corresponds to half the phase
of the internal zitterbewegung motion. 

The Lagrangian is now, instead of (\ref{eqLag}), of the general form
\[
L_0=\dot{t}T+{\dot{\bf r}}\cdot{\bf R}+{\dot{\bf u}}\cdot{\bf U}+{\bomega}\cdot{\bf W}+B\dot{\beta},
\]
Now the conserved quantity under spacetime dilations is 
$
D=Ht-{\bf P}\cdot{\bf r}-B$, 
where $B={\partial L_0}/{\partial\dot{\beta}}$.
If we take the time derivative of this expression and compare with Dirac's equation (\ref{eq:Dir}), 
it results that in the center of mass frame
\[
\frac{dB}{dt}=\pm mc^2,\quad B(t)=B(0)\pm\frac{1}{2}\hbar\omega t.
\]
We nedd the extra variable $\beta$ of the enlarged kinematical group in order to still satisfy
Dirac's equation.
 
\section{The interaction Lagrangian}

The kinematical space of two Dirac particles is spanned by the variables
\[
\{t_a,{\bf r}_a,\beta_a,{\bf u}_a,\balpha_a\}, \quad a=1,2.
\]
We assume that the Lagrangian which describes the compound system is of the form
$L=L_1+L_2+L_I$.

Because the spin is the only intrinsic property of an elementary particle and cannot be modified
by any interaction, the interaction Lagrangian $L_I$ cannot be a function of $\dot{\bf u}_a$
and of $\dot{\balpha}_a$ or equivalently $\bomega_a$. If it is going to be invariant under the local
$SU(2)$ group of local rotations, then it has to be also independent of $\balpha_a$. 
Otherwise the spin definition
of each particle will be modified. If it is also invariant under spacetime dilations, must be independent
of the $\beta_a$.
The spin definition remains the same as in the free case
\[
{\bf S}_a={\bf u}_a\times\frac{\partial L_a}{\partial \dot{\bf u}_a}+\frac{\partial L_a}{\partial{\bomega}_a}={\bf Z}_a+{\bf W}_a,\quad a=1,2.
\]
The interaction Lagrangian will thus be a function of
\[
L_I=L_I(t_a,{\bf r}_a,\dot{t}_a,\dot{\bf r}_a),
\]
and because of (\ref{eq:F}) a homogeneous function of first degree of the derivatives $\dot{t}_a,\dot{\bf r}_a$,
$a=1,2$. If it is going to be invariant under ${\cal W}\otimes SU(2)$, if we call $x^\mu_a\equiv(t_a,{\bf r}_a)$, then we get
\[
L_I=g\sqrt{\frac{\eta_{\mu\nu}\dot{x}^\mu_1\dot{x}^\nu_2}{\eta_{\mu\nu}(x_1^\mu-x_2^\mu)(x_2^\nu-x_1^\nu)}}=
g\sqrt{\frac{\dot{t}_1\dot{t}_2-\dot{\bf r}_1\cdot\dot{\bf r}_2}{({\bf r}_2-{\bf r}_1)^2-(t_2-t_1)^2}}
\]
where $\eta_{\mu\nu}$ is Minkowski's metric tensor and $g$ is a coupling constant with dimensions of action. Incidentaly we can also see
that the Lagrangian is also invariant under the interchange $1\leftrightarrow 2$.

\subsection{Synchronous time description}

Once an inertial observer is fixed it can make a synchronous time description, i.e.
to use as evolution parameter the own observer's time $t$ 
which is the same as the two time variables $t_1$ and $t_2$. In this case
\[
L_I=g\sqrt{\frac{1-{\bf u}_1\cdot{\bf u}_2}{({\bf r}_2-{\bf r}_1)^2}}=g\frac{\sqrt{1-{\bf u}_1\cdot{\bf u}_2}}{r}
\]
where $r=|{\bf r}_1-{\bf r}_2|$ is the instantaneous separation between the corresponding charges
in this frame. We thus have an action at a distance interaction in terms of a single evolution
parameter $\tau$.

An average over the charge position and velocity in the center of mass
of one of the particles 
imply that the interaction becomes the instantaneous
Coulomb interaction, between the center of mass of the first particle 
(which is also the average position of its charge)
and the charge position of the other. 
The average over the other then corresponds to the interaction of 
two spinless point particles when the spin structure is neglected.

It is suggesting that $g\sim \pm e^2$ in terms of the electric charge of each particle. Then the requirement
of invariance under the enlarged ${\cal W}\otimes SO(3)$ group produces a generalization of the
instantaneous electromagnetic interaction between spinning particles.

\subsection{Analysis of a 2-particle system}

The dynamical equation of a free Dirac particle is a fourth-order differential equation for the position of the
charge which can be separated into a system of coupled second order differential equations for the center of mass ${\bf q}$
and center of charge ${\bf r}$ in the form:\cite{dyn}
\[
\ddot{\bf q}=0,\quad \ddot{\bf r}=\frac{1-\dot{\bf q}\cdot\dot{\bf r}}{({\bf q}-{\bf r})^2}({\bf q}-{\bf r})
\]
In the case of interaction the second equation remains the same because it corresponds to the definition
of the center of mass position which is unchanged by the interaction. 
The first equation for particle $a$ is going to be replaced by $d{\bf p}_a/dt={\bf F}_a$ where ${\bf p}_a$
is the corresponding linear momentum of each particle expressed as usual in terms of the center of mass velocity 
\[
{\bf p}_a=\gamma({\dot{\bf q}_a})m\dot{\bf q}_a,\qquad \gamma({\dot{\bf q}_a})=(1-{\dot{\bf q}_a}^2)^{-1/2},
\]
and the force ${\bf F}_a$ is computed from the interaction Lagrangian
\[
{\bf F}_a=\frac{\partial L_I}{\partial {\bf r}_a}-\frac{d}{dt}\left(\frac{\partial L_I}{\partial{\bf u}_a}\right)
\]
For particle $1$ takes the form:
\[
{\bf F}_1=-g\frac{{\bf r}_1-{\bf r}_2}{|{\bf r}_1-{\bf r}_2|^3}\sqrt{1-{\bf u}_1\cdot{\bf u}_2}+
\frac{d}{dt}\left(\frac{g{\bf u}_2}{2|{\bf r}_1-{\bf r}_2|\sqrt{1-{\bf u}_1\cdot{\bf u}_2}}\right)
\]
where there are velocity terms which behave like $1/r^2$ and acceleration terms which go as $1/r$
in terms of the charge separation $r=|{\bf r}_1-{\bf r}_2|$. 

Then the system of second order differential equations to be solved are
\begin{eqnarray}
\ddot{\bf q}_a&=&\frac{\alpha}{\gamma({\dot{\bf q}_a})}\left({\bf F}_a-\dot{\bf q}_a({\bf F}_a\cdot\dot{\bf q}_a)\right)\label{eq:q2}\\
\ddot{\bf r}_a&=&\frac{1-\dot{\bf q}_a\cdot\dot{\bf r}_a}{({\bf q}_a-{\bf r}_a)^2}({\bf q}_a-{\bf r}_a),\quad a=1,2\label{eq:r2}
 \end{eqnarray}
where $\alpha=g/m$ is the fine structure constant in the case of electromagnetic interaction and once all the variables
are dimensionless. 

\begin{figure}
\begin{center}
\includegraphics{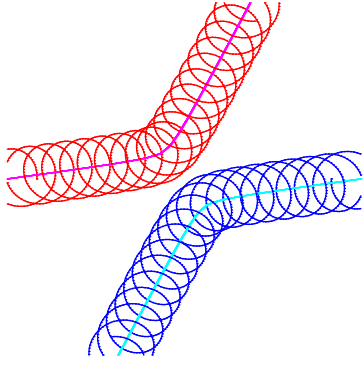}
\caption{The trajectories of the centers of mass and charge of two spinning particles 
of the same charge, with an initial center of mass velocity $v=0.1$ and a small impact parameter.}
\label{fig:scat}\end{center}
\end{figure}

We see in Fig.\ref{fig:scat} the sccatering of two equal charged particles with parallel spins. The trajectory
of each center of mass basically corresponds to the trajectory of a spinless particle coming from the same
initial position as the corresponding center of mass, provided the two particles do not approach each other
below Compton's wavelength. For higher energy proccesses the sccatering of the spinning particles shows
a more detailed structure which depends also on the relative phases of the internal motion of each charge.

\begin{figure}
\begin{center}
\includegraphics{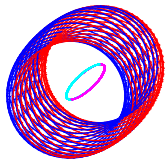}
\caption{Bound motion of the centers of mass and charge of two spinning particles 
of the same charge, 
with parallel spins and with a center of mass velocity $v\le 0.01$, 
for an initial separation between the centers
of masses $0.2\times$Compton's wavelength.}
\label{fig:bound}\end{center}\end{figure}

In Fig.\ref{fig:bound} we also depict the bound motion of the two equal charged particles with parallel
spins. The initial position is below Compton's wavelength. 
The velocity of each particle must be less than $0.01c$ and the phases have to be opposite to each other.
If we try to produce this bound state by a pure collision we need a greater kinetic energy to overcome the
repulsion and the bound state is unstable. But if we think in two conducting electrons in a lattice,
the repulsion is smeared out by the background electrostatic field of the ions and this is a plausible mechanism
for the formation of a bosonic condensate. This feature of formation of metastable bound states
can also be obtained by pure electromagnetic interaction between the two Dirac particles, instead of using
the obtained Lagrangian. Below Compton's wavelength
a repulsion between charges can be transformed into an atraction between the centers of mass.

 \bigskip

 {\small This work has been partially supported by 
Universidad del Pa\'{\i}s Vasco/Euskal Herriko Unibertsitatea grant  9/UPV00172.310-14456/2002.}

 \end{document}